\documentclass[proceedings,proceedingpaper,accept,pdftex,moreauthors]{Definitions/mdpi} 
\firstpage{1} 
\pubvolume{1}
\issuenum{1}
\articlenumber{0}
\pubyear{2025}
\copyrightyear{2025}
\externaleditor{Shin'ichi Nojiri}
\datereceived{1 July 2025} 
\daterevised{16 July 2025} 
\dateaccepted{18 July 2025} 
\datepublished{ } 
\hreflink{https://doi.org/} 

\Title{
New contribution to the anomalous $\pi^0\rightarrow\gamma\gamma$ decay in SU(2) chiral perturbation theory 
}

\TitleCitation{New contribution to the anomalous $\pi^0\rightarrow\gamma\gamma$ decay in SU(2) chiral perturbation theory}

\Author{Zhen-Yan Lu $^{1,2,}$*\orcidA{}, Shu-Peng Wang $^{1}$\orcidB{} and Qi Lu $^{3,4}$\orcidC{}
}

\AuthorNames{Zhen-Yan Lu, Shu-Peng Wang and Qi Lu}

\AuthorCitation{Lu, Z.-Y.; Wang, S.-P.; Lu, Q.}

\address{%
$^{1}$ \quad School of Physics and Electronics, Hunan University of Science and Technology, Xiangtan 411201, China; 
\\
$^{2}$ \quad Hunan Provincial Key Laboratory of Intelligent Sensors and Advanced Sensor Materials, \mbox{Hunan University of Science and Technology,} Xiangtan 411201, China \\
$^{3}$ \quad School of Physics, Beihang University, Beijing 102206, China; luqi0@buaa.edu.cn\\
$^{4}$ \quad CAS Key Laboratory of Theoretical Physics,
            Institute of Theoretical Physics,  Chinese Academy of Sciences,
            Beijing 100190, China \\
}

\corres{Correspondence: luzhenyan@hnust.edu.cn}

\firstnote{Presented at the 5th International Conference on Symmetry, Hangzhou, China, 16–19 May 2025.}

\abstract{
The introduction of axions gives rise to additional one-loop diagrams for the two-photon decays of neutral pions via axion-pion mixing. We compute this correction that has been overlooked in existing calculations, within the framework of SU(2) chiral perturbation theory. Our analysis shows that the correction is proportional to the axion-photon coupling and the square of the axion mass. In the classical axion parameter space, this correction is strongly suppressed by the axion decay constant. However, for QCD axions in the MeV or higher mass range, the correction may become significant. Furthermore, when combined with experimental measurements of the decay width of the $\pi^0 \rightarrow \gamma\gamma$ process, our results rule out the standard QCD axion as a viable explanation for the observed discrepancy between chiral perturbation theory predictions and experimental data.
}


\begin{document}

\section{Introduction} \label{sec:INTRO}

The anomalous decay of the neutral pion~\cite{Bernstein-2011bx,Ananthanarayan-2022wsl} into two photons is a critical process in Quantum Chromodynamics (QCD)~\cite{Gross-2022hyw}, historically confirming the quantum anomaly and the color number $N_c = 3$. This process is described by the Wess-Zumino-Witten (WZW) Lagrangian and serves as a benchmark for testing new physics beyond the Standard Model. The decay width of this process has been extensively studied by using various methods, but the contributions from axion-pion mixing have not been explored in detail~\cite{Lu:2025fke}.

The QCD axion, introduced to address the strong CP problem, shares similarities with the neutral pion and can decay into two photons through its coupling with electromagnetic fields~\cite{85Kaplan215-226NPB,19Alonso-alvarez.Gavela.ea223-223EPJC}. However, its interactions with Standard Model particles are suppressed by the large axion decay constant $f_a$, making it an invisible axion~\cite{DiLuzio-2020wdo,Kim-2008hd,Sikivie-2020zpn}. The axion-photon coupling is central to experimental searches for axions, and recent advancements in SU(2) chiral perturbation theory (CHPT) have enabled precise calculations of this coupling up to next-to-leading order~\cite{GrillidiCortona-2015jxo}. This work aims to analyze the effect of axion-pion mixing on the anomalous decay of the neutral pion to two photons within SU(2) CHPT. By incorporating the WZW Lagrangian, this study provides a high-precision estimate of the mixing contribution. The validity of SU(2) CHPT is supported by its accurate prediction of the topological susceptibility, matching lattice QCD results, namely  \cite{GrillidiCortona-2015jxo,16Borsanyi.others69-71N}
\begin{eqnarray}
\chi_t^{1/4}=
\begin{cases}
75.5(5)~\text{MeV},  \cr
75.6(2)~\text{MeV}  .\cr
\end{cases}
\end{eqnarray}
Moreover, 
 the prediction from CHPT regarding the peak position of the ratio of energy density to the Stefan–Boltzmann limit for strongly interacting matter at finite isospin density is in excellent agreement with results from lattice simulations and Nambu$-$Jona-Lasinio model~\cite{16Carignano.Mammarella.ea51503-51503PRD,Lu-2019diy}. Consequently, we anticipate that this work will help refine our understanding of the interplay between the neutral pion and the QCD axion, with implications for both particle physics and cosmology.

\section{Axion Chiral Perturbation Theory}\label{sec:ChiralLagrangian}

The Lagrangian, including the axion field and the isospin-triplet pion field at leading order (LO) in the chiral expansion, has the form
\begin{eqnarray} \label{eq:LO}
\mathcal{L}_{p^2}
=\frac{F_0^2}{4}\langle D_\mu U^{\dagger}D^\mu U\rangle+
\frac{F_0^2}{4}\langle \chi_a U^{\dagger}+ U\chi_a^{\dagger}\rangle,
\end{eqnarray}
where $\langle \ldots \rangle$ stands for taking the trace in the flavor space, $F_0$ is the pion decay constant in the chiral limit, 
and $\chi_a$ is given by
\begin{eqnarray}  \label{eq:chia}
\chi_a=2 B_{0} \mathcal{M}_{q} e^{i \mathcal{X}_{a}a/f_a}.
\end{eqnarray} 
In the above, $\mathcal{M}_q = \text{diag}\{m_u, m_d\}$ is the diagonal quark mass matrix,  and $B_0 = -\langle \bar{q} q \rangle / F_0^2$ is related to the scalar quark condensate. The axion field is denoted by $a$ and $f_a$ represents the axion decay constant. 
The covariant derivative is given by $D_\mu = \partial_\mu - i [v_\mu, U] - i \{a_\mu, U\}$, where $v_\mu$ and $a_\mu$ are the external vectorial and axial currents, respectively.

The unitary matrix $U(x)$ is expressed as $U(x)=U_0\tilde{U}(x)$, where $\tilde{U}(x)$ parameterizes the chiral fields exponentially
\begin{eqnarray}\label{Uparameter}
\tilde{U}(x)= e^{i\Pi(x)/F_0},~~~
\Pi(x)=\left(
  \begin{array}{cc}
    \pi_3 & \sqrt{2}\pi^+ \\
    \sqrt{2}\pi^- & -\pi_3 \\
  \end{array}
\right),
\end{eqnarray}
where the special unitary matrix $U_0$ represents the ground state, and without loss of generality, it can be chosen to take the diagonal form~\cite{Guo-2015oxa}
\begin{eqnarray}
U_0 = \text{diag}\{e^{i\varphi}, e^{-i\varphi}\},
\end{eqnarray}
where the angle $\varphi$ is determined by minimizing the LO axion potential. It is worth noting that, in general, the symbol $\pi_3$ in Equation~(\ref{Uparameter}) does not correspond to the physical neutral pion. This is because there may be a mixing between the pion and axion fields in the LO Lagrangian, as will be discussed in the following sections.

At next-to-leading order (NLO), only the tree-level term proportional to the coupling $l_7$ contributes to the axion-pion mixing.  
We present here all  tree-level terms
in the $\mathcal{O}(p^4)$ chiral Lagrangian, excluding the derivative terms, namely
\begin{align} \label{eq:l37h13}
 \mathcal{L}^{(4,\text{tree})}= &~ \frac{l_{3}}{16}\left\langle \chi_{a} U^{\dagger}+U\chi_{a}^{\dagger}\right\rangle^{2}
-\frac{l_7}{16}\left\langle\chi_{a} U^{\dagger}-U\chi_{a}^{\dagger}\right\rangle^2 \nonumber\\
& +\frac{h_1+h_3}{4}\langle\chi_{a}\chi_{a}^{\dagger} \rangle+\frac{h_1-h_3}{2}\text{Re}(\text{det}\chi_a),
\end{align}
where the constants $l_7$ and $h_3$ are scale independent, while the constants $l_3$ and $h_1$ contain both ultraviolet finite and divergent parts, which is necessary to eliminate the divergences that appear at higher orders. The latter are
related to the renormalized ones by~\cite{84Gasser.Leutwyler142-142AP}
\begin{eqnarray}
l_{3}=l_{3}^{r}-\frac{\lambda}{2},~~~~~~
h_1=h_1^r+2\lambda, 
\end{eqnarray}
with $\lambda$ the divergence at the space-time dimension $d=4$ in dimensional regularization~\cite{Guo-2015oxa}.

The interaction between axions and photons is conventionally described by the coupling constant $g_{a\gamma\gamma}$, defined by the following Lagrangian 
\begin{eqnarray}
\mathcal{L}_{a \gamma \gamma}=
\frac{1}{8} g_{a \gamma \gamma}
\epsilon^{\mu \nu \rho \sigma} a \mathcal{F}_{\mu v} \mathcal{F}_{\rho \sigma},  
\end{eqnarray}
where $\epsilon^{\mu \nu \rho \sigma}$ is the completely
antisymmetric tensor, $g_{a \gamma \gamma}$ and $\mathcal{F}_{\mu \nu}$ denote axion-photon coupling and the electromagnetic field-strength tensor, respectively. Note that there is some freedom in choosing the diagonal matrix $\mathcal{X}_a$ such that $\langle \mathcal{X}_a \rangle = 1$~\cite{Lu-2020rhp}. If we take $\mathcal{X}_a = \mathcal{M}_q^{-1} / \langle \mathcal{M}_q^{-1} \rangle$~\cite{86Georgi.Kaplan.ea73-78PLB,87Kim1-177PR,Wang-2023xny}, then $U(x) = \tilde{U}(x) = e^{i \Pi(x)/F_0}$, and the LO chiral Lagrangian contains no axion-pion mixing term. In this case, by further incorporating the contribution from the ratio of the electromagnetic and color anomalies, $\mathcal{E}/\mathcal{C}$, the complete $\mathcal{O}(p^4)$ axion-photon coupling can be written as
\begin{eqnarray} \label{eq:garr4full}
g_{a \gamma \gamma}=
\frac{\alpha_{\text{em}}}{2 \pi f_{a}}\left(\frac{\mathcal{E}}{\mathcal{C}}-\frac{2}{3} \frac{4+z}{z+1}\right),
\end{eqnarray}
where $\alpha_{\mathrm{em}} \simeq 1/137$ is the QED fine structure constant, and $z = m_u/m_d$ is the up-to-down quark mass ratio. We stress that in Equation~(\ref{eq:garr4full}), the first term is model-dependent, as it is determined by the specific PQ charges assigned to the quarks. In contrast, the second term is model-independent, emerging from the minimal coupling to QCD at the \mbox{non-perturbative level.}  

\section{Two-Photon Decays of the Neutral Pion}
\label{sec:2rrLagrangianWidth}

The WZW Lagrangian is given by~\cite{01Kaiser76010-76010PRD}
\begin{align} 
\mathcal{L}_{\mathrm{WZW}}^\text{em} =&
 -\frac{N_c}{32 \pi^2} \epsilon^{\mu \nu \rho \sigma}\Big\{
 \langle U^{\dagger} \hat{r}_\mu U \hat{l}_\nu+i \Sigma_\mu(U^{\dagger} \hat{r}_\nu U+\hat{l}_\nu)  \nonumber\\ 
& -\hat{r}_\mu \hat{l}_\nu\rangle \left\langle v_{\rho \sigma}\right\rangle 
+\frac{2}{3}\left\langle\Sigma_\mu \Sigma_\nu \Sigma_\rho\right\rangle\left\langle v_\sigma\right\rangle\Big\},  
\end{align}
where $Q=\mathrm{diag}\{2/3,-1/3\}$ denotes the usual diagonal quark charge matrix for the two-flavor case, and we adopt the following notation
\begin{equation}
\begin{array}{ll}
\hat{r}_\mu=\hat{v}_\mu+\hat{a}_\mu, & \hat{l}_\mu=\hat{v}_\mu-\hat{a}_\mu, \quad \Sigma_\mu=U^{\dagger} \partial_\mu U \\
\hat{X}=X-\frac{1}{2}\langle X\rangle, & \text { for any matrix } X .
\end{array}
\end{equation}
Our goal in this section is to compute the 
two-photon decay width of $\pi^0$ up 
to NLO. 
For the chiral Lagrangian with a minimal set of terms in the anomalous-parity strong sector at $\mathcal{O}(p^6)$, we adopt the notation and Lagrangian provided in Ref.~\cite{02Bijnens.Girlanda.ea539-544EPJC}. Here, only the terms proportional to the coupling constants $c_3^W$, $c_7^W$, $c_8^W$, and $c_{11}^W$ are relevant to the two-photon decays, which are given by
\begin{align}
\begin{aligned}
\mathcal{L}^{(6)}_{\text{ano}}= &~ ic_3^W\epsilon^{\mu \nu \rho\sigma}\left\langle\chi_{-}f_{+\mu \nu} f_{+\rho\sigma}\right\rangle \\
& +ic_7^W\epsilon^{\mu \nu \rho\sigma}\left\langle f_{+\mu \nu}\right\rangle\left\langle f_{+\rho\sigma} \chi_{-}\right\rangle \\
& +ic_8^W\epsilon^{\mu \nu \rho\sigma}\left\langle f_{+\mu \nu}\right\rangle\left\langle f_{+\rho\sigma}\right\rangle\langle\chi_{-}\rangle \\
& 
+c_{11}^W \epsilon^{\mu \nu \rho \sigma}\left\langle f_{+\mu \nu}\right\rangle\left\langle f_{+\gamma \rho} h_{\gamma \sigma}\right\rangle .
\end{aligned}
\end{align}

\begin{figure}[H]
\centering
  \includegraphics[width=0.85\textwidth]{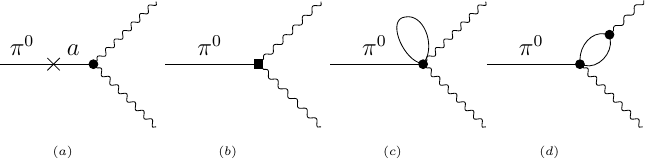}\\
  \caption{ The Feynman diagrams contribute to the anomalous decay of neutral pion: (\textbf{a}) axion-pion mass mixing, (\textbf{b}) $\mathcal{O}(p^6)$ tree-level Lagrangian, (\textbf{c}) the tadpole diagram from WZW Lagrangian, \mbox{(\textbf{c}) the} diagram with one vertex from the WZW Lagrangian and the other from $\mathcal{O}(p^2)$ Lagrangian, and \mbox{(\textbf{d}) the} diagram with both vertexes taken from $\mathcal{L}_{\text {WZW}}$. 
  }\label{fig:diagrams}
\end{figure}

In Figure~\ref{fig:diagrams}, we show the Feynman diagrams contributing to the $\mathcal{O}(p^6)$ corrections to the decay width of $\pi^0 \to \gamma\gamma$. Compared to previous studies, a new non-vanishing $\mathcal{O}(p^6)$ correction arises from the axion-pion mixing, as illustrated in Figure~\ref{fig:diagrams}a. This correction is absent in all prior calculations of the decay width for the $\pi^0 \to \gamma\gamma$ process. 
Let $\delta_i$ denote the relative correction at NLO to the amplitude of the $\pi^0 \to \gamma\gamma$ process. The full decay width is then given by
\begin{align}\label{eq:decaywidthFULL}
\begin{aligned}
\Gamma_{\pi\gamma\gamma} 
=&\Gamma_{\pi\gamma\gamma}^{\text{(LO)}} \bigg\{ 1
 +\underbrace{\frac{16}{9} \frac{m_{\pi}^{2}}{f_{\pi}^{2}}
\bigg[\frac{1-z}{1+z}\left(5 \tilde{c}_{3}^{W}+\tilde{c}_{7}^{W}+2 \tilde{c}_{8}^{W}\right) 
 -3\bigg(\frac{\tilde{c}_{11}^{W}}{4}+\tilde{c}_{3}^{W}+\tilde{c}_{7}^{W}\bigg)\bigg]}_{\delta_\text{tree}}\\
&+\underbrace{2l_7\frac{m_a^2}{f_\pi^2(1+\beta_m)}\frac{1-z}{1+z}
\left(\frac{\mathcal{E}}{\mathcal{C}}-\frac{2}{3} \frac{4+z}{1+z}\right) }_{=\delta_\text{mix}}
\bigg\}^2 ,
\end{aligned}
\end{align}
where $\delta_\text{tree}$ and $\delta_\text{mix}$ represent the corrections arising from the $\mathcal{O}(p^6)$ tree-level Lagrangian~\cite{Ananthanarayan-2002kj,Kampf-2009tk} and 
the contribution from the axion-pion mixing diagram shown 
 in Figure~\ref{fig:diagrams}a, respectively. 
For convenience, we have defined $\tilde{c}_i^W \equiv\left(4 \pi f_\pi\right)^2 c_i^W$, and further replaced the factor $1/f_a^2$ with $m_a^2$ by using the relation between the axion mass and its decay constant, including the NLO correction derived in SU(2) CHPT~\cite{GrillidiCortona-2015jxo}, i.e.,
\begin{eqnarray}\label{eq:mafa}
m_a^2f_a^2=\frac{z}{(z+1)^2}m_\pi^2f_\pi^2(1+\beta_m),
\end{eqnarray}
where $\beta_m$ is given by
\begin{eqnarray}
\beta_m = 
2 \frac{m_\pi^2}{f_\pi^2}\bigg[h_1^r-h_3-l_4^r
+\frac{z^2-6 z+1}{\left(z+1\right)^2} l_7\bigg]
\end{eqnarray}
with the coupling constant $l_4$ replaced by the renormalized one by $l_4=l_4^r+2\lambda$. It is important to note that the replacement can always be done here since the QCD axion mass and its decay constant are uniquely related~\cite{Sikivie-2009qn,Chigusa-2023rrz}. 
And the predicted LO decay width, $\Gamma_{\pi\gamma\gamma}^{\text{(LO)}}$, for $\pi^0\rightarrow\gamma\gamma$ process  
\begin{eqnarray}
\Gamma_{\pi\gamma\gamma}^{\text{(LO)}}
=\frac{\alpha_{\text{em}}^{2}}{(4 \pi)^{3}} \frac{m_{\pi}^{3}}{f_{\pi}^{2}}
=7.763 \pm 0.016~\text{eV},
\end{eqnarray}
which is about 5\% lower than the measurement from the PrimEx-II experiment at JLab, $\Gamma\left(\pi^0 \rightarrow \gamma \gamma\right)=7.802 \pm0.117$~\cite{PrimEx-II-2020jwd}.

From Equation~(\ref{eq:decaywidthFULL}), it is evident that the correction $\delta_{\text{mix}}$ tends to vanish as the axion-photon coupling constant approaches zero. This behavior arises from the cancellation of the anomaly coefficient $\mathcal{E}/\mathcal{C}$ and the QCD correction to the axion-photon coupling constant, as detailed in Equation~(\ref{eq:garr4full}).
In addition, if the axion decay constant is sufficiently large, particularly within the classical axion window of $
10^9~\mathrm{GeV}\lesssim f_a\lesssim10^{12}~\mathrm{GeV}
$~\cite{Kim-2008hd,DiLuzio-2020wdo}, 
the correction $\delta_{\text{mix}}$ may be strongly suppressed. In this case, this contribution would have no noticeable impact on the $\pi^0\rightarrow\gamma\gamma$ process. 

\section{Conclusions}  \label{sec:CONCLUSION}

As an effective field theory of the Standard Model at low energies, CHPT predicts mixing between the QCD axion and the neutral pion. In this work, we investigate the effect of a QCD axion background on the decay width of the neutral pion into two photons within the framework of SU(2) axion CHPT, up to one-loop order. Our results show that the axion background introduces an additional contribution to this anomalous decay via axion-pion mixing. Although this correction is negligible for the standard QCD axion, it is analytically unavoidable. Furthermore, for heavy QCD axions with MeV-scale masses, this new contribution could be significant and comparable to other one-loop effects. 

It should be noted, however, that the calculations in this work are carried out within the framework of SU(2) CHPT, and only the pions and axions are considered. In reality, the $\eta$ and $\eta^\prime$ mesons also exist, and there is a mixing among the $\pi^0$, $\eta$, and $\eta^\prime$ mesons~\cite{16Escribano.Gonzalez-Solis.ea34008-34008PRD,Gao-2022xqz,Gao:2024vkw,Alves:2024dpa,Wang:2024tre}. Therefore, a more realistic physical scenario should take into account the effects of $\pi^0$–$\eta$–$\eta^\prime$ mixing and the contribution from virtual photons, followed by an analysis of their impact on the anomalous decay $\pi^0\rightarrow\gamma\gamma$.

Our results also suggest that similar effects may arise for axion-like particles, which can couple to the neutral pion and have masses in the MeV range or higher, even though they are not directly linked to the strong CP problem. This study highlights the importance of further theoretical and experimental research into axion couplings and their cosmological implications. Additionally, we anticipate future investigations into the dynamics of the QCD axion and axion-like particles, as well as their potential cosmological impacts, such as their behavior in warm and dense media \cite{Zhang-2023lij,Gong:2024cwc}.

\vspace{6pt}


\funding{
This work is supported in part by 
the National Natural Science Foundation of
China (Grant Nos.~12205093, 12405054, and 12375045),  and the Hunan Provincial Natural Science Foundation of China (Grant No.~2021JJ40188).
}



\acknowledgments{The authors would like to thank Feng-Kun Guo, Mao-Jun Yan, Yang Huang and Ji-Gui Cheng for their valuable discussions.
}

\conflictsofinterest{
The author declares no conflict of interest.
} 


\begin{adjustwidth}{-\extralength}{0cm}

\reftitle{References}

\bibliography{RefLuInsp}

\PublishersNote{}
\end{adjustwidth}
\end{document}